\documentclass[runningheads]{llncs}
\usepackage[T1]{fontenc}
\usepackage{graphicx}
\usepackage{amsmath}
\usepackage{booktabs,collcell,eqparbox,makecell}
\usepackage{subcaption}
\usepackage{wrapfig}
\usepackage{multirow}

\usepackage{array}
\usepackage{xcolor}
\newcolumntype{C}[1]{>{\centering\arraybackslash}p{#1}}
\newcommand\MyBox[4]{
\fcolorbox{#1}{#2}{\lower0.75cm
\vbox to 1.7cm{\vfil
\hbox to 1.7cm{\hfil\parbox{1.4cm}{#3\\#4}\hfil}
\vfil}%
}%
}

\def\Plus{\texttt{+}}

\begin{document}
\title{VR-SFT: Reproducing Swinging Flashlight Test in Virtual Reality to Detect Relative Afferent Pupillary Defect}
\titlerunning{VR-SFT: Reproducing Swinging Flashlight Test in Virtual Reality}

\author{Prithul Sarker \and
Nasif Zaman \and
Alireza Tavakkoli
} 

\authorrunning{Sarker et al.}

\institute{Department of Computer Science and Engineering, University of Nevada, Reno, United States \\
\email{prithulsarker@nevada.unr.edu}}

%
%
%
%
\maketitle            
\begin{abstract}
The relative afferent asymmetry between two eyes can be diagnosed using swinging flashlight test, also known as the alternating light test. 
This remains one of the most used clinical tests to this day. 
Despite the swinging flashlight test's straightforward approach, a number of factors can add variability into the clinical methodology and reduce the measurement's validity and reliability. 
This includes small and poorly responsive pupils, dark iris, anisocoria, uneven illumination in both eyes.
Due to these limitations, the true condition of relative afferent asymmetry may create confusion and various observers may quantify the relative afferent pupillary defect differently.
Consequently, the results of the swinging flashlight test are subjective and ambiguous.
In order to eliminate the limitations of traditional swinging flashlight test and introduce objectivity, we propose a novel approach to the swinging flashlight exam, VR-SFT, by making use of virtual reality (VR).
We suggest that the clinical records of the subjects and the results of VR-SFT are comparable. 
In this paper, we describe how we exploit the features of immersive VR experience to create a reliable and objective swinging flashlight test.

\end{abstract}
\section{Introduction}


Relative afferent pupillary defect (RAPD) is a particular eye condition where pupils constrict asymmetrically in response to light stimuli shone in each eye at a time. This condition is also known as Marcus Gunn Pupil. RAPD is caused by a unilateral or asymmetrical retinal or optic nerve disorder. RAPD is not a disease itself, but the defect occurs due to lesions in the afferent pathway which is located behind the pupils. RAPD indicates either of the following: lesion of the optic nerve, optic chiasm, glaucoma, retinal detachment, macular degeneration, dense cataract, amblyopia etc. Even though RAPD is not life threatening in all cases, this usually reveals disease in the pre chiasmal visual pathway~\cite{simakurthy2022marcus}. A strong correlation is found between RAPD and multiple diseases, such as, amblyopia~\cite{miki2008pupillography}, glaucoma~\cite{tatham2014detecting},~\cite{ozeki2013pupillographic},
retinal detachment~\cite{folk1987relative}, 
optic neuritis~\cite{stanley1968swinging},~\cite{cox1981relative}. A study by Wilhelm et al. shows that they noticed asymmetries in the connections between visual pathways and midbrain in about 2\% of normal subjects~\cite{wilhelm2007prevalence}.


The most common eye exam performed to detect RAPD is the swinging flashlight test (SFT)~\cite{simakurthy2022marcus}. 
In normal eyes, the pupils constrict and dilate similarly in response to the light stimuli. If the pupils do not respond identically, they are diagnosed with RAPD.
Neutral density filters (NDFs), cross-polarized filters, and subjective grading based on the degree of initial constriction and subsequent re-dilation of each pupil as the light is swung are different ways for quantifying or assessing RAPDs~\cite{thompson1981measure}\cite{mccormick2002quantifying}\cite{bell1993clinical}. These approaches have been proven to be useful and precise. 
However, human involvement makes these methods highly subjective; so the validity of the measurements is questionable. 
There have been approaches to reduce human dependencies when it comes to measuring RAPD. Most of these strategies tackle this problem using pupillometer~\cite{kawasaki1995variability}\cite{ozeki2013pupillographic}.

Virtual reality (VR) is a programmed environment where the objects and images are made realistic-looking which gives the user the perception of reality, and is one of the most promising technologies in terms of future development. 
In this paper, we introduce a novel method to detect RAPD using VR and discuss in detail about the methodology of our implementation and results. 
By using virtual reality, we eliminate subjectivity of SFT, and control the illumination on each eye precisely to quantify asymmetrical optic nerve disorder.

\section{Literature Review}


The variation in pupillary reactivity under bright and low light environments was originally observed by Marcus Gunn in 1904~\cite{levatin1959pupillary}. 
The disparity in pupillary responsiveness at that time was regarded as peculiar.
Later, Paul Levatin in 1959~\cite{levatin1959pupillary} demonstrated that the pupillary response had a specific cause.
The demonstration was made under the assumption that pupillary impulses or pupillary escape caused by visible light are linearly proportionate to the visual stimuli given.
The authors introduced a novel test to identify pupillary escape under various lighting conditions and optic nerve disease.
This method was named swinging flashlight test (SFT).
However, this approach alone was insufficient to determine the degree of the disease of the optic nerve~\cite{levatin1959pupillary}. 

Keeping the deficits of traditional SFT in mind, Thompson et al. introduced a more reliable technique to measure the relative deficit of the eyes with neutral density filters (NDF)~\cite{thompson1981measure}. The filters dim the light from the flashlight. Based on the reduced light impulse to each eye using the NDF, the authors proposed to quantify the relative afferent defect by the log units of the neutral density filters. The swinging flashlight test is repeated by placing different values of NDF in front of the relatively normal eye until the pupillary reaction is found to same.
In 1993, Bell et al. proposed to quantify RAPD by critically observing the pupillary reaction under different lighting conditions, and identified the period of time for constriction and dilation of the pupil~\cite{bell1993clinical}. The results of this procedure are highly subjective because no measurement tool was utilized. This technique is still the most used one for figuring out RAPD~\cite{bell1993clinical}.

Kawasaki et al. in 1995 used a computerized binocular infrared video pupillometer to record pupillary reaction and experimented with various duration and intensity of light illumination on each eyes~\cite{kawasaki1995variability}. 
The authors found out strong linear relationship between difference in contraction amplitude or pupil diameter (y-axis) and the difference of light illumination for each eye (x-axis). They performed linear regression to locate the point where the line intersects with the x-axis containing the illumination levels. 
The intersection point is the relative afferent pupillary defect score, also known as RAPD score. 
This intersection point is considered because equal and balanced pupillary responses to stimulation of the right and left eyes are produced by the log-unit attenuation.


To make the procedure objective, RAPDx device is introduced~\cite{ozeki2013pupillographic}. 
The device is intended for recording and analyzing pupil responses to various stimuli making quantification of RAPD easy. 
After the recognition of RAPDx device, accurate evaluation of relationship between RAPD and other diseases has been possible. There has been research on patients with optic nerve disease to evaluate progression of RAPD~\cite{satou2016evaluation}.
Satou et al. continued this research and concludes that an absolute RAPD score of 0.2 or less can be considered as healhy whereas an absolute RAPD score of 0.5 log units or over means RAPD is present~\cite{satou2016effects}. 
With the success of deep learning, Temel et al. proposed a transfer learning based approach to capture pupil diameter from video dataset~\cite{temel2019relative}. For the experiment and data collection, the authors developed a headset that can perform automated SFT. From the video, they used a pupil localization algorithm to locate the pupil and measure the pupil diameter, and finally compared left and right pupil diameter to detect RAPD.

Recent improvements in virtual, augmented, and mixed reality (VAMR) have led to widespread acceptance and commercial success which reflect in vision assessment practices. 
Perceptual Modeling in Virtual Reality (PMVR)~\cite{panfili2019effects} and Visual Acuity in Virtual Reality (VAVR)~\cite{zaman2020mixed} conduct their experiments using commercially accessible VAMR technology.
In this paper, we propose a novel objective method to replicate swinging flashlight test using virtual reality for accurate pupil measurement and detect RAPD.
No comparable VR research has been done, as far as we know. 
We propose and validate our approach on two different brands of VR headset. 

\section{Methodology}

In traditional swinging flashlight test, the participant and the examiner are seated side by side in a dimmed room, and the participant is asked to look at a point at a far distance to remove the effects of accommodative response. 
When the examiner thinks that the participant's pupils are completely dilated, a flashlight is directed towards one of the eyes of the participant. 
After the flashlight is turned on, the pupils of a normal participant is constricted. 
Within few seconds of the illumination, the pupils become adjusted to the light illumination. 
Then the flashlight is swiftly moved to the other eye to observe the response of the pupils. 
This process is repeated multiple times, and the examiner takes note of the pupil diameter constriction with response to light and consecutive dilation~\cite{levatin1959pupillary}.
In the case of any participant with RAPD in one eye, when the light stimulus is quickly transferred from the normal eye to the affected eye, the pupils of both eyes dilate~\cite{simakurthy2022marcus}. 

In addition to detecting RAPD, reseachers have worked on measuring the defect of the abnormal eye. 
They found out that the pupillary response to light is difficult to quantify with a reliable number since the amount of constriction varies depending on the pupil's initial size~\cite{thompson1981measure}. 
So, they tried to quantify the relative response rather than quantifying each of them by dimming the stimulus a particular amount until the response in both eyes is equal.

To darken a particular percentage of the stimuli, the most common method is the use of the neutral density filters (NDFs).
When the examiner suspects the relative afferent asymmetry, a NDF with known optical density is placed in front of the suspected eye.
This helps the examiner to easily detect RAPD with different sets of light illumination.
The filters can also aid in decision-making in situations where the afferent deficiency is not too severe ~\cite{thompson1981measure}.
The quantification of the NDF is same as the optical density of the filter and this is measured in log units.
The ratio of the power of the incident beam to that of the outgoing beam is known as optical density (OD).
The equation of the optical density is given by, $OD = \log_{10} (I_O / I_T)$.
Again, fractional transmittance is a measurement of how much optical power gets through the filter, which can be described as, $T = I_T / I_O$.
So, optical density can also be expressed as a function of transmittance, $OD = - \log_{10} T$. 
In the equations, $I_O$, $I_T$ and $T$ are power of the incident light and exiting light and transmittance of the light, respectively.






Instead of reducing light manually, the examiner can hold an NDF in front of the eyes to limit light. 
The NDF quantification starts from optical density of 0.3 and increases with a step of 0.3 log units~\cite{NDF}. 
The details of the NDF quantification is described in table~\ref{table-ndf}. 
By using traditional NDFs, the smallest defect that can be accurately measured is 0.3 log unit.

\subsection{Swinging Flashlight Test in Virtual Reality}

As mentioned earlier, in the traditional swinging flashlight test, the subjects are advised to look at a point at a far distance to remove the effects of accommodation. 
In virtual environment, the screen is very close to the eyes. 
The close screen in VR could easily instigate the accomodative response of the eyes.
To imitate the similar experience of traditional SFT in VR, we used a red "X" which is bigger at first, and with time it reduces itself to a smaller size; making the illusion of the red "X" moving from close to far. 
The subjects are asked to focus on the red "X" at the beginning when the cross is larger in size.
We will mention the red "X" as the target stimuli in the following sections.
With the size of the target stimuli slowly reducing, the subjects gaze at far distance.
This eliminates the effects of the accommodative response for the experiment.
The final size of the target stimuli is very small compared to the view port of the whole screen.
In this way, we make sure that the target stimuli does not have any effect on the pupil dilation and constriction.
After the target stimuli reaches to final size, there is a 5 second delay before any illumination takes place. 
The delay is placed before illumination so that the photoreceptors inside the pupils could compensate for the dark adaptation, and the pupils are properly dilated.



\begin{table}[!t]
  \centering
  \caption{Neutral density filter quantification~\cite{NDF}} \label{table-ndf}
  \begin{tabular}{ C{1.3cm} C{1.4cm} C{2.4cm} C{3.2cm} C{3.4cm} }
    \hline
    \textbf{Optical Density} & 
    \textbf{ND Number} & 
    \textbf{Fractional Transmittance} & 
    \textbf{Luminance in Vive Pro~\cite{VivePro} ($cd/m^2$)} & 
    \textbf{Luminance in FOVE0~\cite{FOVE0} ($cd/m^2$)} \\ 
    \hline
    0.0 &   --      &  100\%   &  97.8  &  97.2 \\
    0.3 &   ND 0.3  &   50\%   &  49.0  &  48.5 \\
    0.6 &   ND 0.6  &   25\%   &  24.5  &  24.2 \\
    \hline
  \end{tabular}
\end{table}

\begin{wrapfigure}{r}{0.6\textwidth}
    \centering
    \includegraphics[width= 0.58\textwidth]{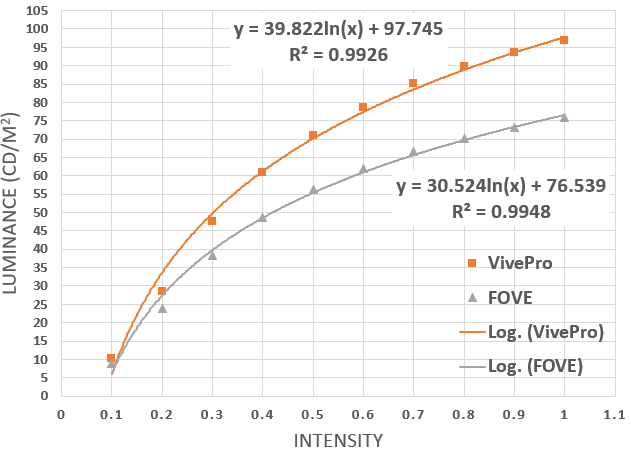}
    \caption{Regression curves of RGB values vs luminance of Vive Pro and FOVE 0.\centering}
    \label{RGBvsLuminance}
\end{wrapfigure}

In the next part, we discuss about how we illuminate each eye in VR. In SFT, only one eye is illuminated at a particular time before quickly moving to the other eye.
In VR headsets of some brands, displaying image in one eye is easier because of their technology.
To illuminate, we show a full white image in each of the eyes. We used same values of red, green and blue (RGB) values so that the resulting color correspond to white. After a particular time delay, we make the screen black on that eye and illuminate the other eye with a full white image. This process is repeated so that the participant gets similar experience of SFT. During the illumination, the target stimuli is always present for both eyes. The participant is asked to focus on that target stimuli during the test.

To replicate the usage of NDF of SFT in VR, we used respective percentage of first RGB values. 
However, luminance of the image shown in the screen does not correlate linearly with the RGB values provided in the VR settings. For example, if RGB values of 1, 1 and 1 are considered as the illumination without any NDF, RGB values of 0.5, 0.5 and 0.5 do not reflect as ND 0.3 or 50\% transmittance. This means even though fractional RGB values are used, the luminance of the screen does not reduce in the same way linearly. Because of this reason, before designing the experiment in VR, we first experimented the luminance of the VR screen using different RGB values. For the measurement, we used i1 display pro~\cite{i1displaypro} device. The device gets the color and luminance of the image at which it is pointed. During the measurement, we input particular RGB values in the respective VR settings, took the reading of the device. The unit of the luminance we get is candela per square meter or $cd/m^2$. 

After obtaining all the data from the devices, we used different regression model to fit the data.
For the models, we used pearson correlation coefficient as the obejective function.
The best regression line is found to be logarithmic lines which has pearson correlation coefficient of over 0.99 for both VR headsets.
The details of the regression curves is shown in fig~\ref{RGBvsLuminance}.
For our experiments, we considered the RGB value of 1 in Vive Pro analogous to the illumination without NDF. From fig~\ref{RGBvsLuminance}, RGB value of 1 for FOVE 0 has less luminance than that of Vive Pro.
For this reason, we used RGB value of 5 in FOVE 0 as the zero optical density.
Based on that value and our regression curve, we measured RGB values of 50\% and 25\% of the illumination which is similar to ND 0.3 and ND 0.6 respectively (table~\ref{table-ndf}).
We used those values to design the experiment.


In classic SFT, the examiner holds the light for about 3 seconds in front of each eye~\cite{thompson1981measure}. 
This duration is referred as pause time~\cite{loewenfeld1993pupil}. 
We designed the experiment in VR in such a way that we could specify the pause time.
We experimented with 0.1, 1, 2 and 3 second pause time.
From those experiments we concluded that with pause time lower than 2 second, the pupils do not get enough time to redilate before the light is again illuminated.
For this reason, we designed our experiments with 2 second and 3 second pause time only.
We kept switching time of illumination between the eyes as 0 second to replicate the procedure of quickly moving to the other eye.
Thus, the midbrain pretectal pupillomotor center can engage with sequential stimuli more readily and has less opportunity to be influenced by supranuclear influences with shorter dark period~\cite{kawasaki1995variability}.


For each of the 2 second and 3 second pause time experiment, we designed 50\% and 25\% illumination in addition to 100\% illumination. 
The order in the experiment is at first we provide 100\% illumination starting from right eye and then, followed by 100\% illumination on left eye.
The next illumination sequences are 50\% and 25\% illumination at right eye, keeping the left eye illumination at 100\%.
In the same way, The following sequences are 50\% and 25\% illumination at left eye, while the right eye illumination is at 100\%.
Each of the sequence is repeated three times.
The tests of 3-second pause time and 2-second pause time are named protocol-1 and protocol-2 of VR-SFT respectively. The total duration of protocol-1 and protocol-2 of VR-SFT is 95 seconds and 65 seconds respectively, including 5-second dark adaptation time at the start of the tests.

\subsection{VR Implementation \& Experimental Software}
The virtual reality application was developed in unreal engine version 4.24. This specific version was selected because both Fove0~\cite{FOVE0} and HTC Vive Pro~\cite{VivePro} Eye devices had plugins that were supported by that engine version. The programs for the VR environment is written in C\Plus\Plus ~programming language. Unreal Engine is used to integrate the programs with the VR headsets which made the incorporation convenient and flexible. We used the FoveHMD\cite{fovePlugin}, and SRanipal\cite{SRanipal} unreal engine plugins to track pupil diameters in the Fove0 and HTC Vive Pro Eye HMD respectively. For accurate eye-tracking data, each subject went through the same automated calibration process before each session.
Unreal engine has a wide support for lighting and illumination that can be configured with real-world units such as $lux$ and $cd/m^2$. However, our experiments require dichoptic light presentation so that stimuli are only visible to one eye at a time. This makes standard illumination sources unviable as they would be visible to both eyes. Therefore, custom material shaders were created to render in either left or right displays of VR HMDs. As these shaders needed to self-illuminate, they were configured by interpolating the corresponding emissive value from the RGB vs light intensity graph from fig~\ref{RGBvsLuminance}.

\begin{figure}[!t]
    \centering
    \includegraphics[width= 0.9\textwidth]{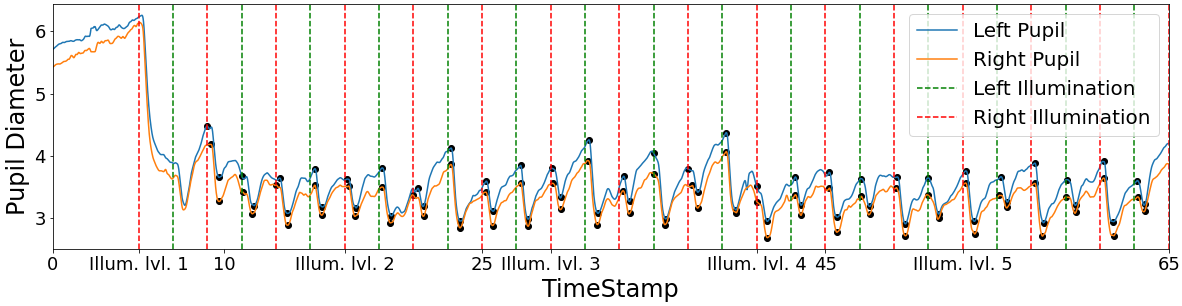}
    \caption{Timestamp vs pupil diameter for 2 second protocol (protocol-2). The maximum and minimum pupil diameters in a particular illumination for each pupil are marked as black dot points on the signal. The respective maximum and minimum values are used to calculate the CA for any particular illumination. \centering}
    \label{Fig-PupilDiameterSignal}
\end{figure}

In this experiment, we needed to ensure that light sources turned on and off with minimal temporal errors. The error is generally not minimal when the event tick functionality in unreal engine is used. Instead, we opted to use time delegates for illumination start, illumination end and a separate pause between mini-sessions. This flexible configuration allowed us to compare different pupillary reflexes to variation in time duration. To induce the effect of a relaxed pupil at the start of a test session, a plane (75cm$\times$75cm) with a red "X" in the middle was gradually moved to a distance of 100m from 2m. After a 5s initial delay, the emissive material corresponding to the left eye starts to illuminate.

\subsection{RAPD Scoring}

The method we use in this paper to quantify asymmetric pupillary responses is RAPD score~\cite{cohen2015novel}. The authors of~\cite{cohen2015novel} recorded pupil responses to SFT using RAPDx Pupillometer introduced by Konan Medical USA. Before we describe RAPD score, we need to get familiar with a term called constriction amplitude (CA). CA is defined as the difference between maximum diameter before illumination and minimum diameter after illumination, given in eq.~\ref{eq-percentage-change}. The computation of maximum and minimum pupil diameter is shown in fig~\ref{Fig-PupilDiameterSignal}. In our analysis, we take the average CA of direct and consensual pupillary responses to account for the contraction anisocoria (unequal pupil sizes), same as~\cite{cohen2015novel}. 

\begin{equation}
    \label{eq-percentage-change}
    \% \: \text{Change in CA} = \frac{\text{Max diameter - Min diameter}}{\text{Max diameter}}
\end{equation}

Now, the ratio of percentage change in CA to right and left eye illumination is calculated as RAPD score. 
Based on the presumption that an afferent defect will alter the pupil reflex, the eq.~\ref{eq-RAPDx} was developed. The authors also assumed that the effective reduction in perceived light intensity is inversely correlated with pupillary reflex~\cite{cohen2015novel}. The unit of RAPD score is log units.

\begin{equation}
    \label{eq-RAPDx}
    \text{RAPD score} = 10\log_{10} (\frac{\text{Right CA change during right eye illumination}}{\text{Left CA change during left eye illumination}})
\end{equation}

\begin{wrapfigure}{r}{0.5\textwidth}
    \centering
    \includegraphics[width= 0.48\textwidth]{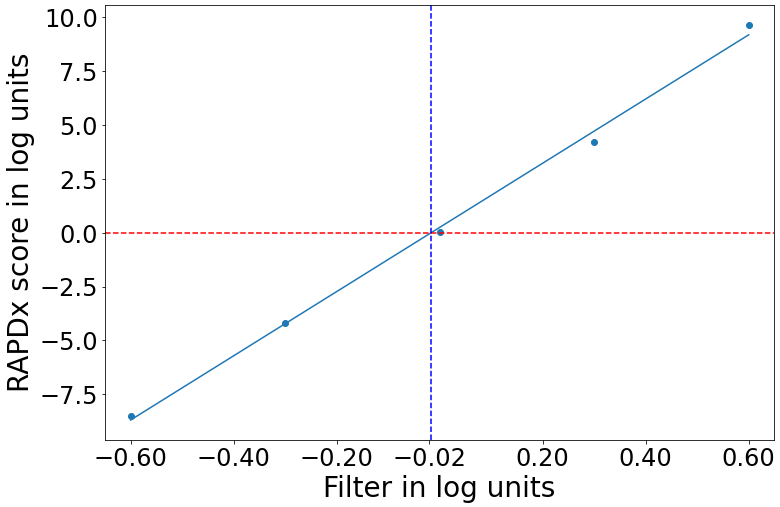}
    \caption{RAPD score calculation. The RAPD scores for different optical density are plotted as points and a linear regression model is used to find the optimum straight line. The straight line intersects x-axis at a particular point which is the final RAPD score. \centering}
    \label{Fig-RAPDx}
\end{wrapfigure}

To get a final RAPD score based on the all the illumination levels, we first calculate the RAPD score of individual illumination levels, and get 5 RAPD scores for 5 different illumination levels. Subsequently, we plot RAPD scores at y-axis with respect to respective illumination levels at x-axis. Both RAPD score and illumination level units are log unit.
Finally, the point at which the best fitted straight line obtained using linear regression model intersects x-axis is the RAPD score of that particular participant. In this way, rather than depending on RAPD score for only one illumination level, we evaluate and take into consideration the RAPD scores of other illumination levels. 
The disadvantages of taking only the mean value of the RAPD scores over linear regression method is if any participant blinks a lot or does not focus on the target stimuli, there could be deviation from the original RAPD score. This is why linear regression ensures those outliers do not interfere with the final RAPD score. Visual descriptions of the process is given in fig~\ref{Fig-RAPDx}.

\section{Dataset}

After the tests are performed, the data is exported to a comma separated values (csv) file. 
The data were kept in a secured folder in the cloud. Each of the participant data file has 5 columns. The columns include timestamp, right eye illumination level, left eye illumination level, right pupil diameter, and left pupil diameter.
The research study was approved by the Institutional Review Board (IRB).
The inclusion criteria for the controls were: age between 18 to 85 and no history of RAPD while the inclusion criteria for the patients were: age between 18 to 85 with clinical records of RAPD. All of the patients were referred to this study by their respective eye doctors. In total, 36 controls and 4 RAPD positive patients participated in the study. 
Out of 36 controls, the number of female and male participants is 8 and 28 respectively. 
On the other hand, the number of female and male RAPD positive patients is 3 and 1 respectively.
The mean age of the controls and RAPD positive patients are 26.4 years and 65.3 years.

\section{Data Analysis \& Results}

For our analysis, each data file was considered at a time. 
As both protocol-1 and protocol-2 take over 60 seconds, the participants blink during the tests.
For detection of blink, we used a sliding window method and eliminated those particular rows of the file. 
Moreover, to remove noise, we used a Gaussian filter with 6 standard deviation for the Gaussian kernel to the pupil diameter data. 
Next, we separate the filtered values into individual illumination level and their respective pause time for each of the eyes. 
From there, we find the maximum and minimum (fig~\ref{Fig-PupilDiameterSignal}) to calculate the CA using eq.~\ref{eq-percentage-change}. 
As each of the illumination level is repeated 3 times for each eye, we compute the change in CA for all repetitions.
To minimize the effect of anisocoria and negate the effect of the outliers, the changes in CA for particular eyes are averaged.
The change in CA of right eye and left eye for each of the illumination level is then used to calculate the RAPD score as stated in eq.~\ref{eq-RAPDx}.
The ratio of change in CA to right and left eye stimulation is calculated as RAPD score.
The RAPD score represents the asymmetry of pupillary reaction to the stimulation of any particular eye.

\begin{figure}[!t]
    \centering
    \begin{subfigure}[h]{0.48\linewidth}
        \centering
        \includegraphics[width=0.8\linewidth]{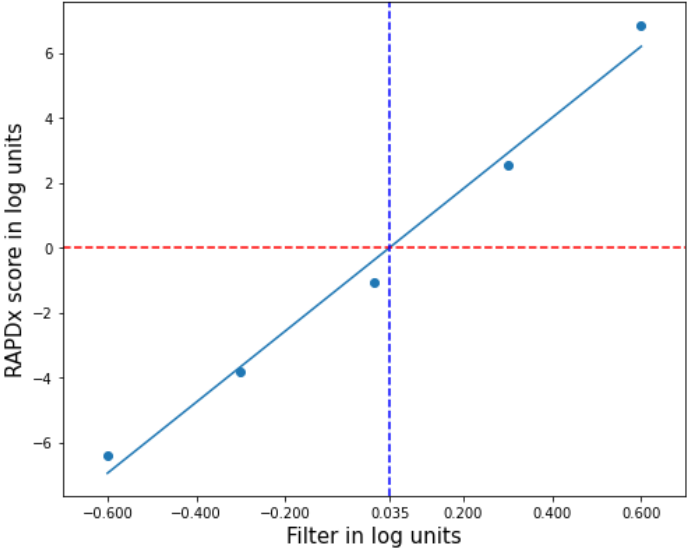}
        \caption{RAPD negative data}
    \end{subfigure}
    \hfill
    \begin{subfigure}[h]{0.48\linewidth}
        \centering
        \includegraphics[width=0.8\linewidth]{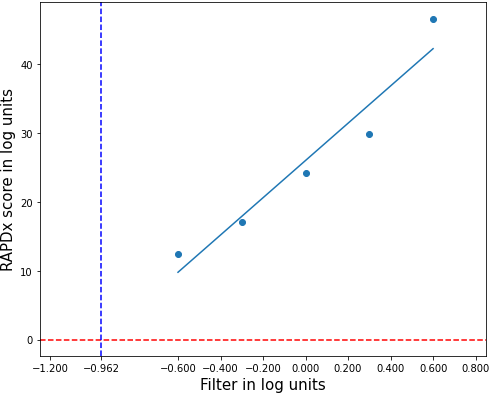}
        \caption{RAPD positive data}
    \end{subfigure}%
    \caption{The left figure represents a RAPD negative result as the intersection point is between -0.3 to 0.3. However, in the right figure, the result is of a left eye RAPD positive participant because the intersection point is at -0.96 log units, which is over the threshold.\centering}
    \label{fig-rapd-control-patient}
\end{figure}


Since we have 5 different illumination levels, we obtain 5 RAPD scores for each of them.
In the next step, we plot the RAPD scores against their respective illumination levels in log units (fig~\ref{Fig-RAPDx}), and use a linear regression model to find the best fitted straight line as stated in the methodology section.
To achieve the final RAPD score, we find the intersection point of the straight line and the x-axis.
The intersection point is the final RAPD score.
We use the threshold of $\pm0.3$ to separate between RAPD positive and RAPD negative similar to~\cite{kawasaki1996long}.
If the absolute value of the final RAPD score is less than 0.3, the participant is considered as RAPD negative; while the absolute value is over 0.3, the participant is considered as RAPD positive. Based on the location of the intersection point, we can figure out which eye has RAPD. If the intersection point is at below -0.3, RAPD is present in the left eye. On the other hand, if the intersection point is over 0.3, RAPD is present in the right eye of that participant. (fig~\ref{fig-rapd-control-patient})

The performance metrics used in this paper are accuracy, sensitivity, and specificity. 
Accuracy = $\frac{1}{N} \sum (TP + TN)$, Sensitivity = $\frac{TP}{TP + FN}$, and Specificity = $\frac{TN}{TN + FP}$ where $TP$, $FN$, $TN$ and $FP$ are true positive, false negative, true negative and false positive respectively.
The results from both protocol-1 and protocol-2 are same. 
For both protocols, VR-SFT misdiagnosed one case only. One control participants was found positive, and the final RAPD score was 0.49. The individual has a history of lasik surgery which could lead to asymmetric pupillary responses. To confirm, we have requested the person to have an eye check-up. 
The accuracy, sensitivity, and specificity of both of the protocols of VR-SFT are 97.5\%, 100\%, and 97.22\% respectively.
The findings demonstrate that VR-SFT was successful in identifying every positive case.








\section{Discussion and Future Work}

The outcomes clearly illustrate that our use of virtual reality headset yields the same outcomes as clinical results.
We presented identical results in terms of RAPD scores for two brands of virtual reality headsets, which supports our theory and detection system.
With approaches from the literature, this confirms the validity of our method.
Additionally, our findings demonstrate strong linear correlation between RAPD scores and the corresponding light illumination levels.


Some of the participants have reported after the study that they were able to see the target stimuli twice, when the light illumination changes between the eyes. 
We hypothesize that this occurs as a result of each eye's residual light illumination. 
After any illumination is turned on, the participants have a residual vision of target stimuli from its dark interval. 
At that precise moment, the other eye's light illumination is switched off, allowing the participant to view the target stimuli with that eye. The target stimuli for both eyes do not overlap since the subject sees two of them with each eye. The subject, therefore, appears to be experiencing double vision.
Moreover, VR sickness is a major setback for using any VR headset. The possibility of VR sickness exists even though the subjects did not move during our experiment. Three of the participants from control group have reported symptoms of VR sickness after completion of the experiment. The triggered sickness could potentially have effect on the diameter of the pupils. Again, any element from the external environment such as noise during the test, could impair pupil dilation and constriction as described in~\cite{zekveld2018pupil}.
These reasons could cause the difference between the points and the best fitted straight line for calculating the RAPD scores.

In this paper, we demonstrated a novel approach using virtual reality to detect a particular eye condition, called relative afferent pupillary defect (RAPD).
We hypothesize that the strategy we used should have a significant influence on the detection of RAPD as well as other eye related illnesses.
With the advancement of VR headsets, the pupil detection and frame rate are improving day by day.
Again, virtual reality headsets are very low-weight and only require one cable to connect to the computer, making the devices more portable.
Eventually, patients from all around the world  will have access to install the RAPD detection program of VR-SFT to any commercially available VR.
In future, the patients will be able to perform the routine tests by themselves from their home.
We hope to continue this investigation and thoroughly improve the methods in our upcoming works.
This includes limiting the duration of the test, and superior method to remove the effects of accommodation.
As we followed the measure employed in earlier studies in this work, we would like to have a better metric in our future work that can be utilized to classify data more efficiently.

\subsubsection{Acknowledgements} 
Portions of this material is based upon work supported by the Office of the Under Secretary of Defense for Research and Engineering under award number FA9550-21-1-0207.


\bibliographystyle{ieeetr}
\bibliography{bibliography.bib}

\end{document}